\documentclass[journal]{IEEEtran}
\usepackage{graphicx} 
\usepackage{url}
\usepackage{hyperref}
\usepackage{booktabs}

\title{Integrated Heterogeneous Service Provisioning: Unifying Beyond-Communication Capabilities with MDMA in 6G and Future Wireless Networks}

\author
{

Pengyi~Jia,~\IEEEmembership{Member,~IEEE},
Xianbin~Wang,~\IEEEmembership{Fellow,~IEEE},
Yongxu~Zhu,~\IEEEmembership{Senior Member,~IEEE},
Shi~Jin,~\IEEEmembership{Fellow,~IEEE}, and
Robert~Schober,~\IEEEmembership{Fellow,~IEEE}

\thanks{Pengyi Jia and Xianbin Wang (corresponding author) are with Western University, Canada; Yongxu Zhu and Shi Jin are with Southeast University, China; Robert Schober is with Friedrich-Alexander University Erlangen-Nuremberg, Germany.}

}

\begin{document}

\maketitle
\begin{abstract}
The rapid evolution and convergence of wireless technologies and vertical applications have fundamentally reshaped our lifestyles and industries. Future wireless networks, especially 6G, are poised to support a wide range of applications enabled by heterogeneous services, leveraging both traditional connectivity-centric functions and emerging beyond-communication capabilities, particularly localization, sensing, and synchronization. However, integrating these new capabilities into a unified 6G paradigm presents significant challenges. This article provides an in-depth analysis of these technical challenges for integrative 6G design and proposes three strategies for concurrent heterogeneous service provisioning, with the aggregated goal of maximizing integration gains while minimizing service provisioning overhead. First, we adopt multi-dimensional multiple access (MDMA) as an inclusive enabling platform to flexibly integrate various capabilities by shared access to multi-dimensional radio resources. Next, we propose value-oriented heterogeneous service provisioning to maximize the integration gain through situation-aware MDMA. To enhance scalability, we optimize control and user planes by eliminating redundant control information and enabling service-oriented prioritization. Finally, we evaluate the proposed framework with a case study on integrated synchronization and communication, demonstrating its potential for concurrent heterogeneous service provisioning.
\end{abstract}

\vspace{-0.4cm}
\section{Introduction}
\vspace{-0.15cm}
The unprecedented advancements in communications and computing technologies have revolutionized our lifestyles, industrial practices, and societal structures. With the ongoing expansion and paradigm shift of consumer and industry applications, future wireless networks (e.g., 6G) are expected to support a multitude of complex applications through seamless collaboration among interconnected devices and machines \cite{Road6G}. Emerging advanced applications like multi-sensory extended reality (XR) and intelligent Internet of Everything (IoE) will require future wireless networks to concurrently satisfy diverse requirements of multiple heterogeneous services, including communications, sensing, localization, synchronization, and collaborative computing.

\vspace{-0.4cm}
\subsection{The Need for Heterogeneous Services with Beyond-Communication Capabilities}
\vspace{-0.1cm}

To meet these diverse requirements, 6G and future networks must incorporate new capabilities for tailored \textit{heterogeneous service provisioning}. Unlike connectivity-centric 5G networks, which categorize communication scenarios primarily by usage needs, 6G must address the diverse service demands of advanced applications by leveraging new capabilities such as precise positioning and support of artificial intelligence (AI), as envisioned in the IMT-2030 objectives \cite{A4}. These \textit{beyond-communication capabilities}, based on their enabling signals, can be broadly summarized into two categories:

\textbf{Sensing-centric new capabilities}, such as localization, target detection, and tracking, are achieved through the dual use of wireless signals to probe the radio propagation environment between transmitters and targets. These capabilities depend on temporary collaboration, where the distributed user equipment (UE), base stations (BS), and the external environment engage in short-term coordination to accomplish task-specific sensing objectives. Through these coordinated interactions, 6G systems can accurately measure propagation-related signal attributes, enabling real-time environmental awareness and system adaptation capability to changing conditions. These capabilities become fundamental to support situation-aware decision-making for advanced services in future networks.

\textbf{Communication-augmented new capabilities} like precise time synchronization, collaborative computation, and network-enabled AI, represent a significant evolution from traditional communication functions. Unlike conventional approaches focused on reliable data transmission, these advanced capabilities involve dynamic exchanges of non-conventional data and sophisticated protocols that enable enhanced collaboration among multiple connected devices. By optimizing dynamic network topology while accounting for latent relationships, long-term collaborations among network entities can be facilitated to support application-oriented system design. This helps develop new network orchestration strategies essential for the adaptability and intelligence of future networks.
\begin{figure*}
    \centering
    \includegraphics[width=0.68\linewidth]{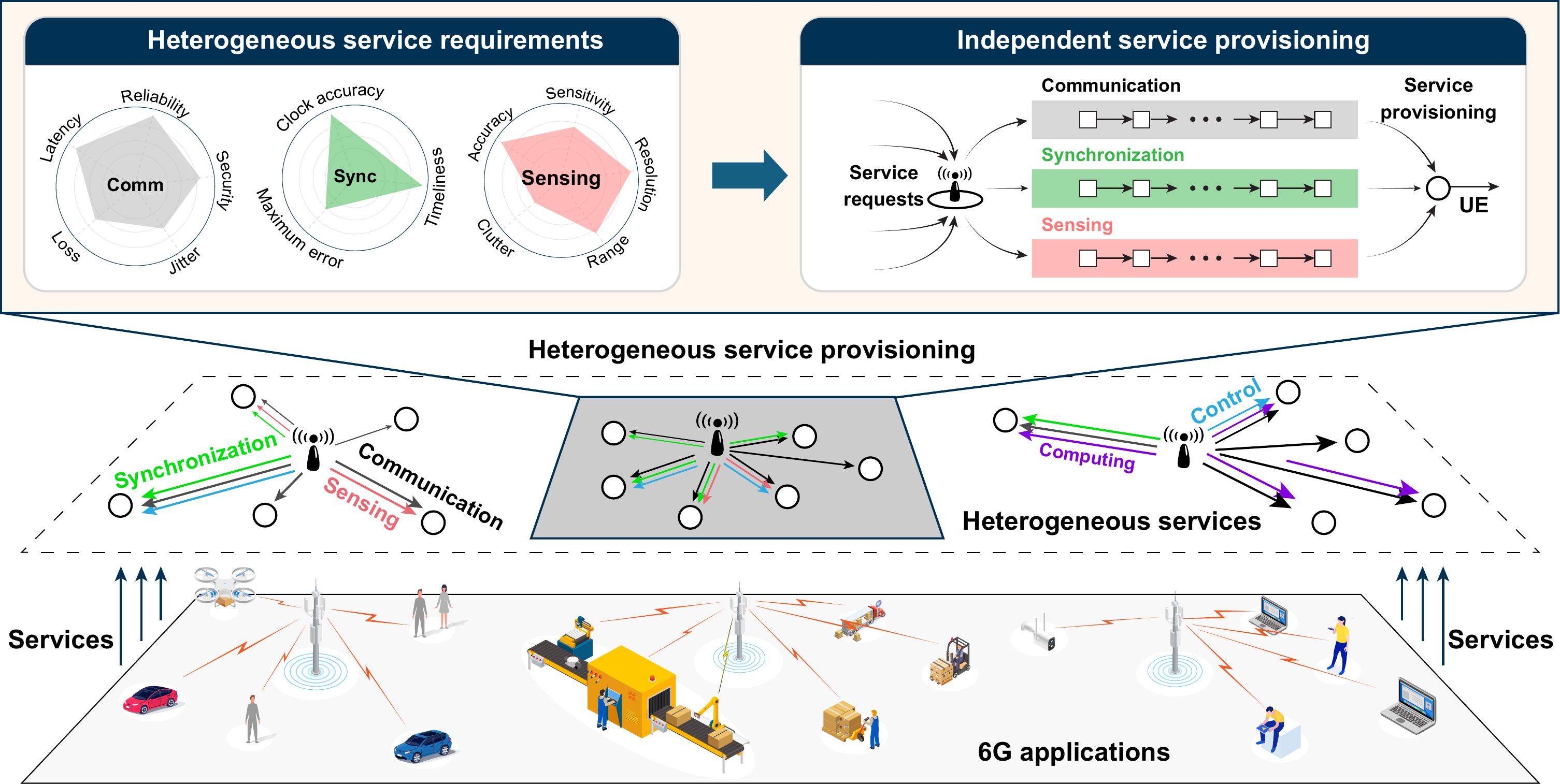}
    \caption{Challenges in traditional connectivity-centric network design, such as the heterogeneous service requirements of diverse applications and the separated design of beyond-communication capabilities, hinder the service provisioning for advanced applications in the 6G era.}
    \label{fig:1}
\end{figure*}

\vspace{-0.4cm}
\subsection{Key Challenges of Integrated Service Provisioning}
\vspace{-0.1cm}

Compared to traditional connectivity-centric functions, the complex vertical applications enabled by 6G and future networks require simultaneous creation and orchestration of multiple new capabilities to support heterogeneous services. This need has driven converged designs such as integrated sensing and communication (ISAC) \cite{Robert}. However, as service diversity and system scale continue to grow, existing connectivity-centric network architectures become increasingly inefficient, as illustrated in Fig. \ref{fig:1}. This paradigm shift towards heterogeneous service provisioning brings the following challenges.

\subsubsection{Lack of Heterogeneous Service Integration Mechanisms}

Current network protocols and architectures provision different services through separate processes, including service requests, link/environment calibration, service initialization, and provisioning. This separation hinders the shared use of limited resources across heterogeneous services, leading to underutilized resources such as communication infrastructure, radio spectrum, protocol stack, and situational knowledge (e.g., channel states, timestamps, and user data). This inevitably results in redundant signal processing and substantial resource waste. For instance, timestamp information could be utilized for multiple services like time synchronization, precise localization, and data alignment \cite{DataAlignment}. Nonetheless, it is frequently generated and transmitted redundantly due to the isolated design of these capabilities. Consequently, this inefficiency undermines the network's ability to effectively support heterogeneous services, especially with resource constraints.

The rigid and single-purpose design of current networks, focused primarily on data transmission, further exacerbates the challenges of simultaneously supporting multiple services with limited radio resources. In conventional connectivity-centric designs, resource allocation is centered on a narrow set of data transmission use cases, which can result in wasted resource utilization for communication and insufficient resources for new capabilities. Given the explosive growth in the number of UEs (e.g., IoE devices) with heterogeneous service requests, a dramatic increase in service density within a given spectrum is anticipated. Even with the envisaged advancements in 6G technologies, such as Terahertz communication and holographic radio, the available resources could still be insufficient for this multi-user multi-service environment, where significant multipath effects and inter-capability interference will challenge the achievable service quality for future applications.

\subsubsection{Ineffective Orchestration for Integrated Heterogeneous Service Provisioning}

Existing 5G and even initial 6G research often relies on scenario-specific resource allocation, which fails to address the heterogeneous demands of emerging applications. These networks categorize services into predefined types such as eMBB, uRLLC, and mMTC in 5G, as well as ubiquitous connectivity and ISAC in 6G. However, the coexistence of sensing-centric and communication-augmented beyond-communication capabilities, coupled with the arbitrary multi-dimensional service demands, introduces increasing heterogeneity that surpasses these rigid patterns. Applying uniform orchestration methods across varied applications risks significant wastage of limited communication resources.

The challenge is further compounded by the variability of service demands, which current network designs are ill-equipped to handle. The mobility of UEs and the opportunistic deployment of small BSs lead to non-uniform service demand distributions across the network. Additionally, dynamic channel conditions and periodic fluctuations driven by user activities result in time-varying service satisfaction. These spatial-temporal variations necessitate more adaptable and responsive orchestration strategies. However, existing scenario-specific designs lack the flexibility to dynamically adjust to these evolving conditions, leading to inefficient resource utilization and compromised service quality.

Moreover, traditional orchestration strategies often prioritize the optimization of each capability for individualized service provisioning. This narrow focus inevitably overlooks the potential synergies and interdependencies between coexisting capabilities. For example, accurate synchronization can enhance time-of-flight measurements crucial for high-precision localization, while precise localization data can aid synchronization during coordination processes. However, current orchestration methods typically assess the quality of each service in isolation, failing to capture the holistic performance of multi-service applications. Addressing this issue requires new orchestration strategies that emphasize the mutual enhancement of different capabilities, unlocking the synergistic value that improves overall service quality.

\subsubsection{Dramatically Increased Complexity for Heterogeneous Service Provisioning}
Efficient network management and resource utilization in 6G and future networks are closely tied to the design of the control and user planes. With the expansion of service variety and network scale, the control and feedback signaling required in future networks for heterogeneous service provisioning will grow exponentially. Additionally, the protocol headers associated with various service data could become overwhelming, especially those for communication-augmented new capabilities. This issue is aggravated when small payloads are transmitted for multiple services, resulting in substantial overhead on both user and control planes.

In existing connectivity-centric designs, each service for every UE requires its own set of control processes, such as signaling, scheduling, and QoS management, along with separate service data transmission. In a multi-user multi-service scenario, the cumulative overhead from these separated designs can escalate dramatically. This excessive overhead results in degraded service provisioning performance, manifesting as increased latency and reduced overall efficiency in 6G and future networks.

Moreover, the complexity of resource allocation and scheduling in this multi-user multi-service environment introduces significant optimization challenges. Real-time monitoring of user and application dynamics, combined with the need to adaptively adjust resource allocation strategies, significantly increases global optimization complexity. Centralized processing and optimization methods further exacerbate this issue given the limited processing power in each local UE and stringent service timeliness requirements. This increasing complexity can lead to delays and suboptimal solutions, significantly impacting the performance of multi-service applications. Therefore, efficiently managing both user and control plane resources, while optimizing the concurrent support of multiple services, becomes crucial for 6G network design.

\vspace{-0.8cm}
\subsection{Scope of This Article}
\vspace{-0.2cm}
As discussed above, the existing connectivity-centric network paradigm cannot efficiently support concurrent heterogeneous services anticipated in future applications. The creation, orchestration, and control of diverse capabilities must be seamlessly unified with traditional connectivity functions through new network frameworks. To enable efficient heterogeneous service provisioning, this article focuses on answering the following critical questions:

\vspace{-0.1cm}
\begin{itemize}
    \item How can multiple new capabilities be unified within a shared network infrastructure and protocol to concurrently enable heterogeneous service provisioning?
    \item How can the increasingly complex yet limited communication resources be optimized for tailored heterogeneous services in a specific application scenario, and how can resources, processes, and information of these capabilities be leveraged to maximize the integration gain?
    \item How can the complexity and overhead of integrated heterogeneous service provisioning be balanced dynamically with service quality to ensure better scalability?
\end{itemize}



\vspace{-0.1cm}
The remainder of this article is organized as follows. First, we propose a unified 6G paradigm based on multi-dimensional multiple access (MDMA) to address these issues, efficiently unifying beyond-communication capabilities for integrated heterogeneous service provisioning. As an example, we then apply this unified paradigm to support integrated synchronization and communication services. Several future directions are discussed before drawing conclusions.

\vspace{-0.3cm}
\section{Unifying Beyond-Communication Capabilities for Integrated Service Provisioning}

\begin{figure*}
    \centering
    \includegraphics[width=0.701\linewidth]{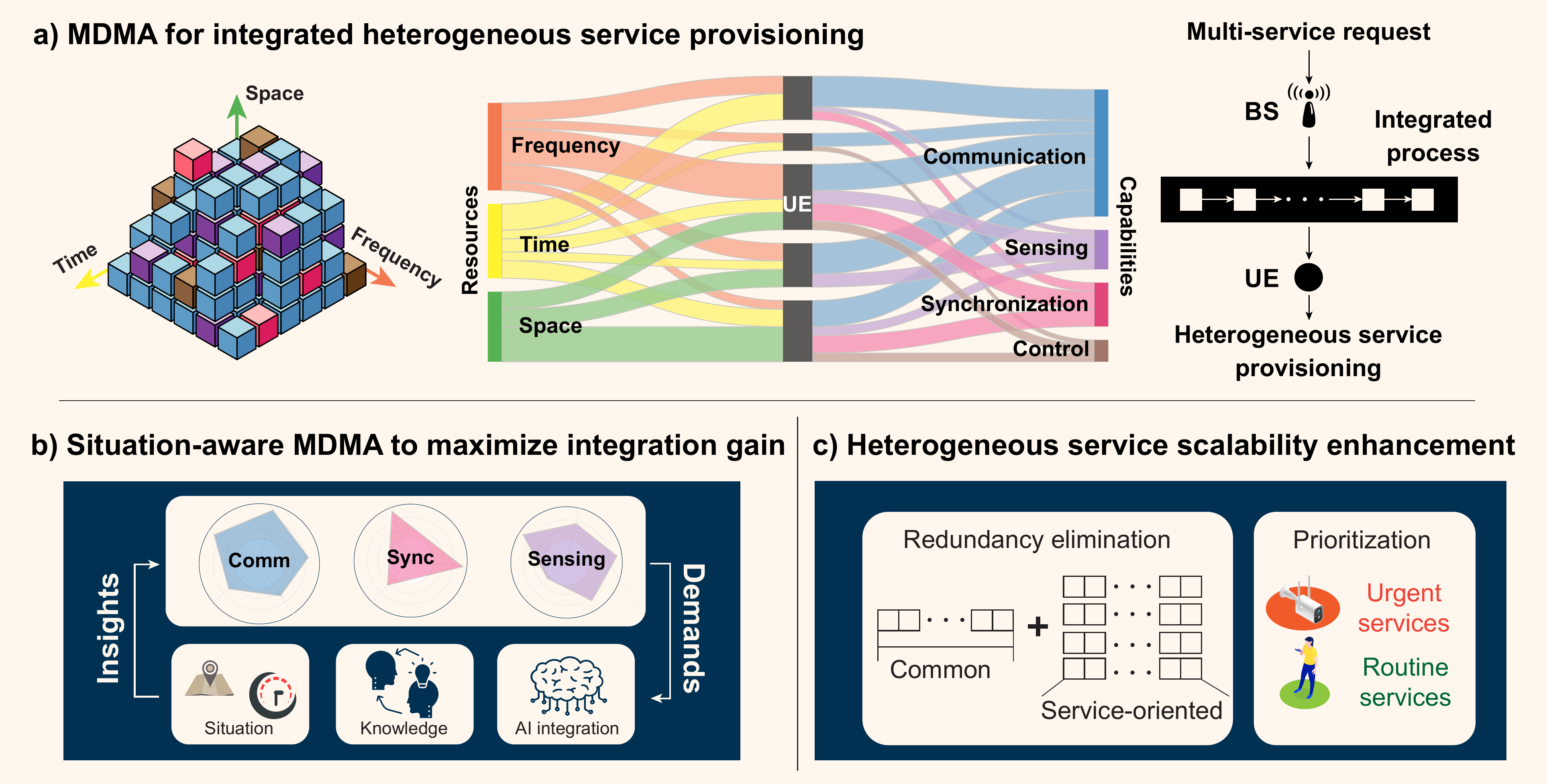}
    \caption{The proposed unified paradigm for integrated heterogeneous service provisioning with MDMA-based service integration and orchestration. }
    \label{fig:2}
\end{figure*}

To address the challenges inherent to existing connectivity-centric designs, a unified paradigm is proposed for 6G and future networks to support integrated heterogeneous service provisioning. As illustrated in Fig. \ref{fig:2}, the proposed MDMA-based platform effectively unifies multiple capabilities with shared resource utilization, flexible heterogeneous service orchestration, and scalable service provisioning.

\vspace{-0.3cm}
\subsection{MDMA for Integrated Heterogeneous Service Provisioning}
\vspace{-0.08cm}

Supporting diverse applications through integrated heterogeneous service provisioning necessitates the tight coordination of multiple new capabilities due to their shared access to radio resources and mutual impact. To achieve this, we propose a unified framework using MDMA as a flexible platform for concurrently enabling communication-centric and beyond-communication capabilities with coordinated resource sharing.

Achieving this relies on flexible shared access either orthogonally or non-orthogonally and coordinated management of all available radio resources in different dimensions. The proposed MDMA framework can unify and flexibly allocate multi-dimensional resources (e.g., time, space, frequency) for supporting concurrent capabilities, ensuring efficient resource sharing among heterogeneous capabilities within a unified platform \cite{wang2023realizing}. In this regard, either a communication-centric or a beyond-communication process is generally considered as an ``access" request to multi-dimensional radio resources.

MDMA also expands resource allocation to all possible dimensions for optimal resource utilization. It opportunistically assigns available resources based on current network conditions and device capabilities, enabling seamless coordination that minimizes mutual interference among non-orthogonal users \cite{MDMA1}.  To effectively support heterogeneous service provisioning, MDMA must adopt a service-centric framework that dynamically maps service-specific access requests to multi-dimensional resource blocks in real time, allowing the network to create and customize new capabilities in response to evolving heterogeneous service needs. Moreover, service-centric resource multiplexing and partitioning enable the efficient coexistence of heterogeneous new capabilities across multiple users by accounting for the criticality of services. This coordinated approach minimizes inter-capability interference during integrated heterogeneous service provisioning, laying a robust foundation for supporting more complex scenarios.

\vspace{-0.4cm}
\subsection{Situation-Aware MDMA for Integration Gain Maximization}
\vspace{-0.1cm}

The integrated design and information sharing among concurrent heterogeneous capabilities offer significant mutual benefits, which can be evaluated in terms of \textit{integration gain}. This gain quantifies system-wide improvements in service quality and resource efficiency through optimized inter-service coordination and dynamic resource allocation. MDMA enables integration gain through multi-purpose resource utilization, where each resource block could concurrently support multiple services, particularly those engaged in supporting the same application. For instance, a single resource block can simultaneously support communication and sensing services, leveraging their correlated demands to enhance resource efficiency and service quality. Additionally, service data from different communication-augmented capabilities can be flexibly combined by aligning their application-specific service requirements. This adaptive approach enables MDMA to optimize resource allocation dynamically, allowing limited resources to support a broader range of services.

This gain can be further amplified by incorporating situational awareness from beyond-communication capabilities. Spatial-temporal insights derived from sensing, positioning, and synchronization allow MDMA to interpret dynamic network conditions, such as mobility patterns, traffic variations, and interference levels \cite{MDMA2}. By leveraging these insights, MDMA can dynamically adjust the sharing of resource blocks among multiple services, ensuring optimal alignment with real-time network demands. This situation-aware orchestration minimizes inter-service conflicts during integration with improved resource efficiency. Furthermore, AI-driven predictive models improve adaptability by anticipating future service requirements and resource availability \cite{MARL}. By proactively optimizing the orchestration of heterogeneous services, MDMA maximizes service integration gain, ensuring resource allocation aligns seamlessly with future system conditions.

\vspace{-0.4cm}
\subsection{Enhancing Service Scalability and Effectiveness via MDMA Optimization and Resource Prioritization}
\vspace{-0.1cm}
As the number of users and services in 6G networks continues to grow, ensuring the scalability of integrated service provisioning is critical. In addition to the growing service data traffic, the frequent scheduling and real-time feedback required by MDMA-based service orchestration result in significant control overhead, complicating the management of heterogeneous services. The control and user plane separation architecture, a key enabler for flexible resource management, must be further refined to handle the increasing complexity of services. This involves enhancing the control plane for efficient signaling and optimizing the user plane to support higher traffic capacities.

\vspace{0.1cm}
\textbf{Optimizing MDMA Overhead for Scalable Service Provisioning}: Integrating various capabilities across different protocol layers enables the multi-purpose reuse of network architecture and information, significantly reducing overhead in both control and user planes. In the control plane, overhead stems from managing and coordinating the resource allocation and scheduling in MDMA to support heterogeneous services. Unifying control and feedback signaling at the physical layer is crucial for simultaneously managing multiple capabilities, enabling scalable service coordination with reduced signaling \cite{OverheadRed}. In the user plane, focusing on transmitting essential service data minimizes the overhead caused by static or redundant information, thereby lowering transmission volume and processing load. Techniques such as header compression and service data unit aggregation \cite{HeaderCompression} effectively eliminate redundancies (e.g., static routing information, identical packet headers, and unchanged timestamp digits) while supporting multiple heterogeneous services. Therefore, by addressing superfluous control information and redundant data transmissions, the proposed MDMA-based platform can scale efficiently to accommodate a broader range of heterogeneous services.

\vspace{-1cm}
\textbf{Control Resource Prioritization for Effective Service Orchestration:} While the separation of control and user planes enhances management efficiency, these planes can be further customized to perform complementary functions that improve network adaptability and service effectiveness. For example, Data over Non-Access Stratum (DoNAS \cite{tan2022breaking}) enables infrequent user service data transmission via signaling messages, leveraging the control plane for lightweight data transfers with enhanced service quality. However, the limited resources of the control plane constrain the number of services it can support efficiently. To address this issue, prioritizing control plane resources becomes crucial, particularly for high-value services requiring greater reliability and timeliness. Such services should be granted preferential access to control plane resources for essential service data transmission. By contrast, lower-priority services can be orchestrated through more strategic mechanisms, such as clustered control among adjacent nodes or piggybacking control information with user-plane data. This joint optimization alleviates the burden on the control plane in MDMA-based design, ensuring the effectiveness of critical services even under stringent resource limitations.

\section{Case Study: Integrated Synchronization and Communication Service Provisioning}

Concurrently enabling synchronization and communication is essential for mission-critical industrial applications, which rely on precise timing and reliable data exchange within local area access networks. Traditional synchronization methods, like Precision Time Protocol (PTP), need specialized hardware and packets to exchange timestamps and estimate time deviations. The frequent interactions can disrupt ongoing data transmission and significantly consume control plane resources, thereby affecting overall network performance \cite{R1}.

To overcome these limitations, this case study designs an MDMA-based integrated synchronization and communication (ISynC) framework that unifies these capabilities within the same PHY and MAC layers. By leveraging shared network resources, situational information reuse, and efficient control overhead management, the framework maximizes integration gains while meeting the unique requirements of different services. Through this case study, we validate the feasibility of integrating heterogeneous services with enhanced provisioning quality and scalability.
\vspace{-0.4cm}
\subsection{Service-oriented MDMA for ISynC Capability Coexistence}
\vspace{-0.1cm}
Integrated synchronization and communication service provisioning requires meeting the heterogeneous quality of service (QoS) requirements of both capabilities. Synchronization services emphasize precision and timeliness as two key metrics to ensure reliable timing service delivery. Precision reflects the local time error relative to the reference, while timeliness measures the delay between a synchronization service request and its fulfillment. In contrast, communication capabilities focus on traditional connectivity-centric QoS metrics such as latency and throughput, commonly optimized through signal quality-based management. Due to the maturity of these aspects, further details on communication capabilities are omitted.

To accommodate these heterogeneous service demands, we adopt service-oriented MDMA as the integrative platform for ISynC, enabling the mapping of multi-dimensional radio resources to service access requests. Synchronization services with stringent precision and timeliness requirements are allocated additional time-domain resources to ensure prompt and reliable timestamp delivery. Meanwhile, communication services are optimized through allocations across other resource dimensions (e.g., frequency and spatial resources) to minimize interference and maximize throughput. This service-oriented multiplexing ensures the coexistence of synchronization and communication capabilities with the desired performance.

The integration gain of ISynC is maximized through optimized resource utilization and inter-service mutual enhancement. Synchronization service data can be opportunistically integrated with communication processes to meet heterogeneous service requirements. For example, synchronization data may piggyback on communication packets during low-urgency periods, allowing both services to share radio resources temporarily. This eliminates the need for separate synchronization operations, reducing control overhead and improving resource efficiency. By dynamically adjusting service access, MDMA flexibly accommodates multiple services while maintaining acceptable performance. Additionally, ISynC leverages the mutual enhancement of synchronization and communication capabilities. Reliable communication ensures accurate delivery of synchronization timestamps with improved precision and timeliness. In turn, precise synchronization enhances situational awareness, enabling real-time optimization of MDMA resource allocations under dynamic network conditions. These integration gains ensure that ISynC efficiently meets the complex demands of heterogeneous service provisioning.

\vspace{-0.4cm}
\subsection{ISynC Implementation with Existing Network Frameworks}
\vspace{-0.1cm}
Integrating timestamps into existing communication architectures requires balanced transmission time interval (TTI) allocation between communication and synchronization services to support service-oriented MDMA design. In the 5G NR MAC architecture \cite{R19}, data and control signaling are encapsulated within protocol data units (PDUs), which are further divided into sub-PDUs to support multiple functions. This flexible structure allows synchronization data to coexist with communication data in the same PDU. The frame structure of a downlink MAC PDU, shown in Fig. \ref{fig:3}a), demonstrates how sub-PDUs can carry service data units (SDUs) for user data or control elements (CEs) for MAC layer signaling. The allocation of TTI directly affects the size and scheduling frequency of these PDUs, impacting how SDUs and CEs are packaged and transmitted. These considerations open two possible ISynC schemes:

\begin{figure}
    \centering
    \includegraphics[width=0.91\linewidth]{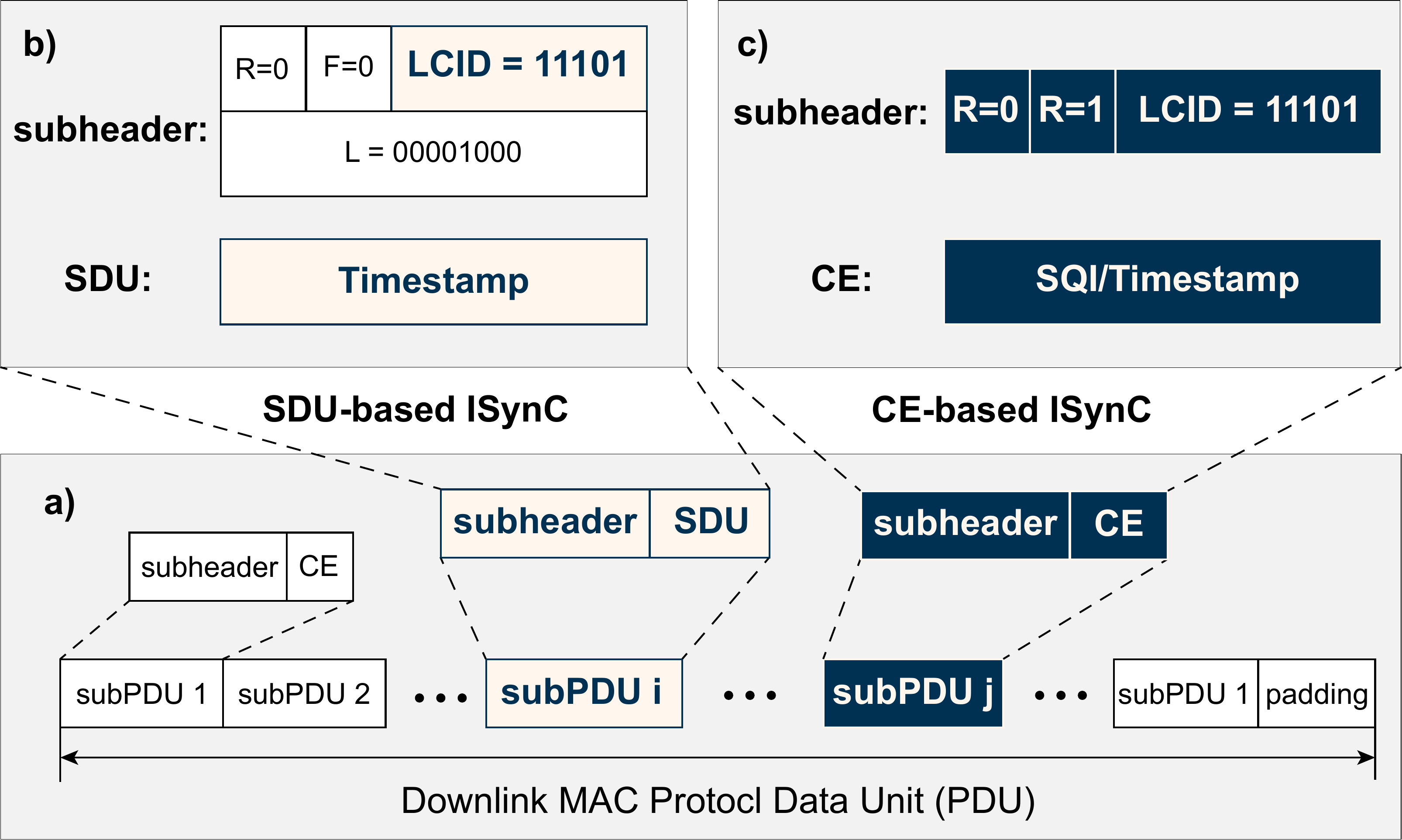}
    \caption{The 5G NR MAC frame structure with synchronization capability integrated via SDU-based and CE-based ISynC design.}
    \label{fig:3}
\end{figure}

\textbf{SDU-based ISynC}: It is intuitive to integrate synchronization service data using SDUs designated for user data. As shown in Fig. \ref{fig:3}b), each SDU can include synchronization service data such as timestamps, feedback, and synchronization quality indicators (SQI). The subheader provides critical information, including the logical channel identifier (e.g., LCID=29) and the length of service messages (e.g., L=8) for proper coordination. Although SDU-based ISynC integrates synchronization and communication with minimal complexity, it may become inefficient when a larger number of UEs require frequent synchronization services. Allocating an entire SDU for transmitting small payloads (e.g., an 8-byte timestamp) for each UE introduces notable overhead in the user plane. Furthermore, sharing the user plane with other services can lead to resource competition, affecting the quality of ISynC services due to the competition within the same TTIs.

\textbf{New CE-based ISynC}: An alternative approach is to design a new CE specifically for ISynC data exchange. This method transmits small-sized timestamps and SQIs via the control plane as CEs instead of using SDUs in the user plane. This can enhance synchronization accuracy and timeliness due to prioritized processing in different protocol layers, which also allows for more efficient use of TTIs. As illustrated in Fig. \ref{fig:3}c), ISynC service data are encapsulated within the CE after reserving a specific logical channel, with two reserved bits used to indicate data types. While the CE-based ISynC design offers greater flexibility, it also introduces additional control plane overhead that requires careful management.

\vspace{-0.4cm}
\subsection{Cluster-based Hybrid ISynC for Scalability Enhancement}

The SDU-based and CE-based ISynC schemes face challenges of service quality degradation and limited control plane resources. To meet the heterogeneous demands of synchronization and communication services while enhancing scalability in large-scale systems, we propose a service-oriented hybrid framework that jointly optimizes these two schemes.

\begin{figure}
    \centering
    \includegraphics[width=0.92\linewidth]{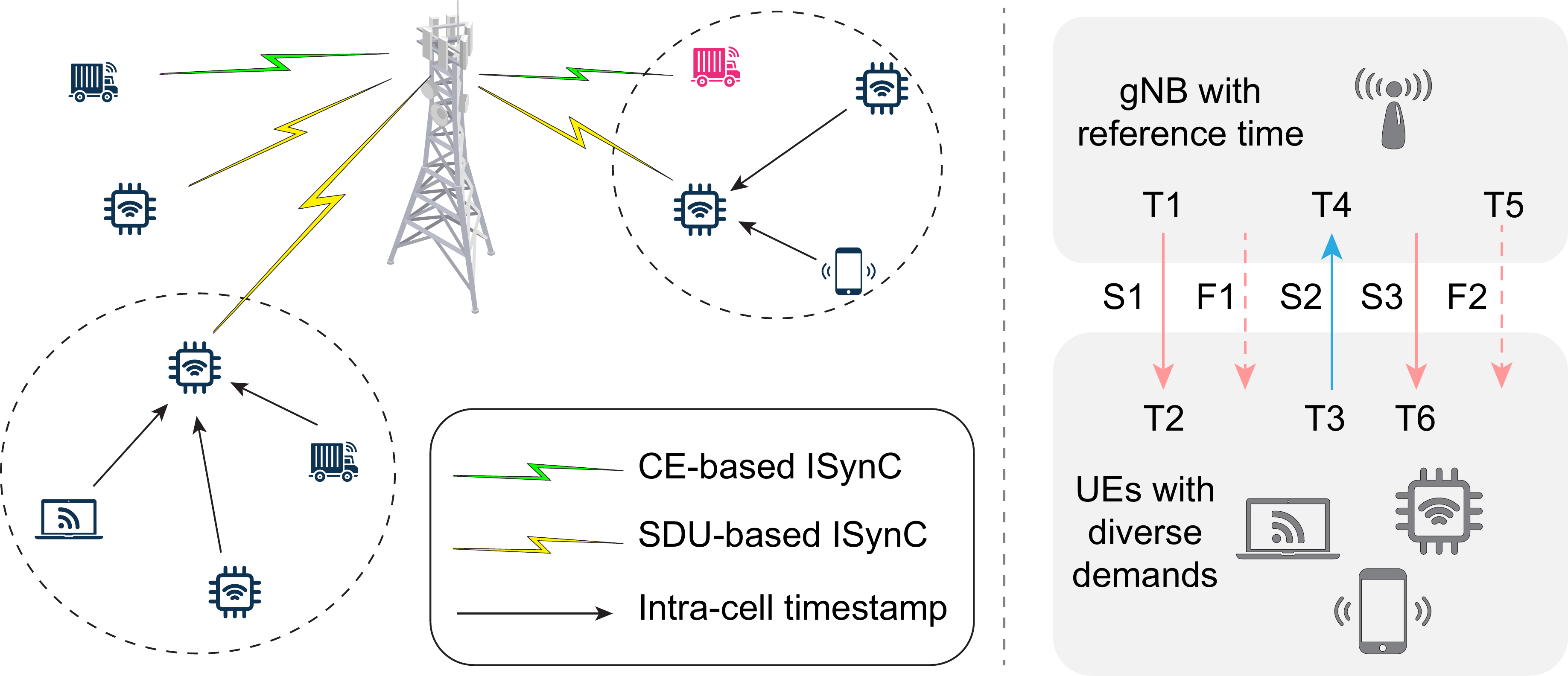}
    \caption{Clustered ISynC design based on the service urgency for distributed UEs in a local area access network with service prioritization.}
    \label{fig:4}
\end{figure}

As depicted in Fig. \ref{fig:4}, UEs in the system are categorized based on the value of their services, which is evaluated using weighted service requirements for synchronization and communication. To prevent overloading the control plane, CE-based ISynC is selectively applied to high-value UEs with stringent demands. For other UEs, synchronization messages are transmitted using SDUs, which can be inefficient due to excessive packet headers associated with small payloads. To mitigate this, we introduce a clustered ISynC framework, aggregating multiple SDUs from various UEs. Non-prioritized UEs are organized based on their real-time locations, with a cluster head designated to collect and process synchronization service data. This aggregation reduces redundant packet headers from cluster members by transmitting a single larger SDU, which improves efficiency and reduces user plane overhead. By prioritizing control plane resources for high-value UEs and aggregating non-prioritized service data in the user plane, the hybrid ISynC framework enhances overall network performance and service scalability.

Moreover, timestamp-based synchronization relies on frequent timestamp exchanges to estimate clock parameters (e.g., PTP estimates clock offset using four packet transmissions). In the proposed ISynC, clock skew and offset are estimated using six timestamps, as shown in Fig. \ref{fig:4}. This process begins with an initial synchronization flag (S1), followed by two timestamped packets (S2 and S3), and may include optional follow-up messages (F1 and F2) if physical layer timestamping is used to enhance accuracy.

Table \ref{tab:1} details the timestamp transmission process, highlighting the asymmetric nature of the exchanges, where BSs send multiple timestamps to UEs to estimate local clock parameters with reduced uplink burden. The uplink is only used to send the SQI, which determines synchronization frequency. To minimize service data redundancy, timestamp compression is applied, retaining only the time information that differs from previous synchronization exchanges. This can significantly reduce packet size based on the synchronization frequency.

\vspace{-0.4cm}
\subsection{Performance Evaluation}
\vspace{-0.1cm}
In this section, we present numerical simulations to evaluate the performance of the proposed ISynC framework in a 5G NR network, comparing it with traditional separated service provisioning methods. This evaluation focuses on service integration gains and overhead reduction.

\begin{figure}
    \centering
    \includegraphics[width=0.76\linewidth]{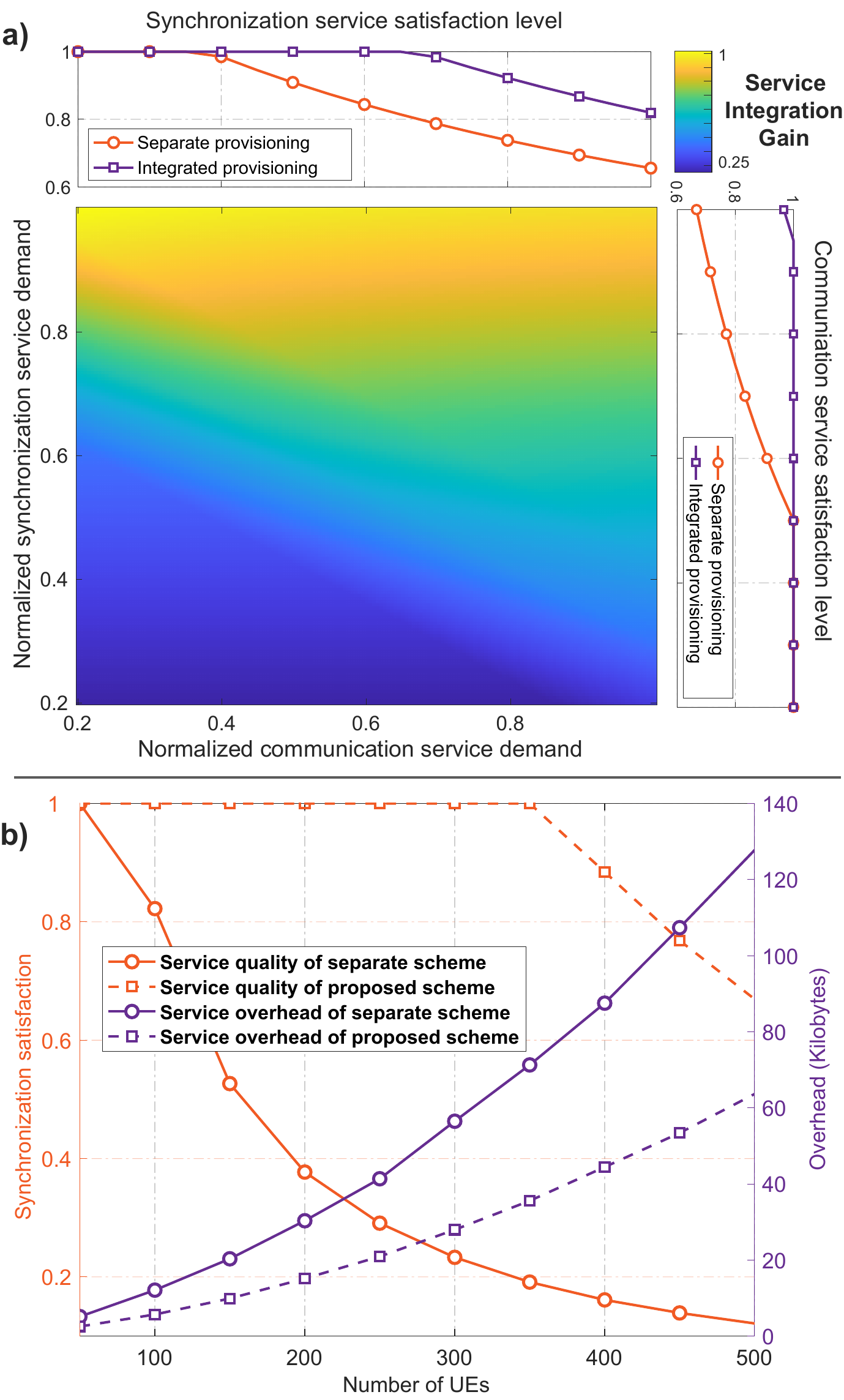}
    \caption{Performance of ISynC: a) Service satisfaction level and service integration gain. b) service quality improvement for different network scales.}
    \label{fig:5}
\end{figure}

As shown in Fig. \ref{fig:5}a), the simulations consider varying service requirements for synchronization and communication. The results indicate that as the demand for both services increases, the service satisfaction rate of traditional methods drops dramatically. Conversely, ISynC maintains a consistently higher satisfaction rate, even under stringent conditions. Notably, ISynC fully meets communication service demands across various scenarios, underscoring its robust handling of increasing synchronization requirements. The service integration gain heat map shows significant improvement with ISynC, particularly under high synchronization demands, validating the mutual benefits of integrating synchronization within the communication framework.
\begin{table}[t]
    \centering
    \caption{The timestamp message transmission between BS and UE}
    \begin{tabular}{c c c c}
    \toprule
        Message & R bits & Content & Info available at UE \\
    \midrule
        S1 & DL-00 & Sync flag & T2 \\
        F1 & DL-01 & T1 & T1, T2 \\
        S2 & UL-00 & SQI & T1 -- T3 \\
        S3 & DL-10 & Compressed T4 & T1 -- T4, T6 \\
        F2 & DL-11 & Compressed T5 & T1 -- T6 \\
    \bottomrule
    \end{tabular}
    \label{tab:1}
\end{table}

We also analyze ISynC's performance across different network scales, ranging from 50 to 500 UEs. Figure \ref{fig:5}b) shows that traditional methods suffer from reduced synchronization satisfaction and increased overhead as the number of UEs rises. In contrast, ISynC efficiently supports more UEs, maintaining full service satisfaction while reducing overhead by more than 50\%. These results highlight ISynC's capability to enable more efficient heterogeneous service provisioning, especially in complex and large-scale network environments.

\vspace{-0.3cm}
\section{Future Directions: Value-oriented Heterogeneous Service Integration}

While the foundational integration enables multiple capabilities to coexist and share resources efficiently, future networks must evolve to become intelligent and value-oriented to maximize the integration gain of heterogeneous services.

\textbf{Holistic Context Sharing for Intelligent Heterogeneous Service Integration}: The heterogeneous service demands and dynamic resource availability in 6G necessitate intelligent service provisioning for real-time optimization and adaptability. Current network designs lack dynamic context to orchestrate heterogeneous services by precisely matching heterogeneous service provisioning and application requirements in resource-constrained environments. Holistic context sharing can vertically integrate system status data across network layers (e.g., network statistics and user behavior \cite{R16}) to comprehensively capture system dynamics, and horizontally share it among services to enable context-aware orchestration. This direction allows MDMA to intelligently satisfy diverse service demands with limited resources, making it essential for mission-critical applications like intelligent manufacturing.

\textbf{Value-oriented Heterogeneous Service Integration based on Knowledge Extraction}:  Evaluating and predicting the value of heterogeneous service provisioning in 6G systems is increasingly challenging, particularly under resource constraints that require trade-offs between competing services. This challenge is especially critical in complex application scenarios such as autonomous transportation, where communication services must seamlessly integrate with environmental sensing to ensure safety and efficiency. Data from prior service provisioning offers valuable insights for value-oriented integration \cite{IntegrationGain}. By analyzing historical performance data, usage patterns, and application requirements, knowledge extraction helps quantify the perceived value of integrating heterogeneous services, reflecting how different combinations influence specific application outcomes. Such insights can guide resource allocation strategies that adapt dynamically to future demands, enabling MDMA to optimize integrated service orchestration and maximize system performance.

\vspace{-0.3cm}
\section{Conclusions}
\vspace{-0.1cm}
This article first overviewed the challenges of integrating beyond-communication capabilities into 6G and future networks. We then outlined a unified 6G network paradigm based on MDMA for integrated heterogeneous service provisioning by optimizing resource utilization and service orchestration. By adopting service-oriented MDMA for multi-capability coexistence, maximizing integration gains through situation-aware service provisioning, and optimizing control and user planes to enhance scalability, we demonstrated how 6G and future networks can evolve to a beyond-communication paradigm. A case study on an integrated synchronization and communication framework showcased the practical advantages of this approach, emphasizing its potential to concurrently support diverse services with heterogeneous demands. Finally, future directions are presented in achieving value-oriented heterogeneous service integration.

\vspace{-0.3cm}
\bibliographystyle{IEEEtran}
\bibliography{main.bib}










\end{document}